\documentstyle[twoside,fleqn,espcrc2]{article}
\input epsf

\newcommand{\AmS}{{\protect\the\textfont2
  A\kern-.1667em\lower.5ex\hbox{M}\kern-.125emS}}

\title{Heatbath Noise Methods in Lattice QCD}

\author{Walter Wilcox\address{Department of Physics, Baylor 
								University, Waco, TX 76798-7316}}
       
\begin{document}

\begin{abstract}
In a recent paper, de Forcrand has pointed out that
matrix inversions using Gaussian noise need not be iterated to
full convergence, but instead may be solved
approximately and treated as a heatbath. Gaussian noise
however is not optimal for minimizing variance. It shown here how his algorithm
may be generalized to a mixture of Gaussian and Z(N) noise,
resulting in a smaller effective variance for some operators.
 
\end{abstract}

\maketitle

\section{Introduction}

Usually, one solves $A\vec{\phi} = \vec{\eta}$, where $\vec{\eta}$ is a noise vector,
to full convergence. DeForcrand pointed out\cite{phil} that this is not
necessary for Gaussian noise. Formally,
\begin{eqnarray}
<{\cal O} > = \frac{1}{Z_{\phi}} \int D \vec{\phi}\,\, e^{-|A\vec{\phi}|^{2}}
{\cal O}.
\end{eqnarray}
Introduce an auxilliary field,
\begin{eqnarray}
<{\cal O} > = \frac{1}{\pi^{N} Z_{\phi}} \int D \vec{\chi}\int D \vec{\phi}
\, e^{- |A\vec{\phi}|^{2} - |\vec{\chi}-A\vec{\phi}|^{2}} {\cal O}.
\end{eqnarray}
Consider a change,
\begin{eqnarray}
\vec{\phi} \rightarrow \vec{\phi}'; \vec{\phi}' = \vec{\zeta} - \vec{\phi},
A\vec{\zeta} = \vec{\chi} - \vec{r}, \vec{\chi} = A\vec{\phi} + \vec{\eta},
\end{eqnarray}
where $\vec{\eta}$ is a complex Gaussian noise vector and $\vec{r}$ is the residual vector
in the solution for $\vec{\phi}'$. One can show the change is accepted with probability
\begin{eqnarray}
P_{A}(\vec{\phi} \rightarrow \vec{\phi}') = {\rm min}(1,e^{-\Delta S}),
\end{eqnarray}
where
\begin{eqnarray}
\Delta S &=& |A\vec{\phi}'|^{2} +|\vec{\chi} - A\vec{\phi}'|^{2} -
|A\vec{\phi}|^{2} - |\vec{\chi} - A\vec{\phi} |^{2}, \nonumber \\
&=& 2 {\rm Re} (\vec{r}^{\dagger} \cdot A(\vec{\phi}-\vec{\phi}')).
\end{eqnarray}

With the assumption that $A\vec{\phi}$, $A\vec{\phi}'$, and $\vec{r}$ are uncorrelated
Gaussian vectors of variance $N$, $N$ and $\epsilon^{2}N$ respectively ($N$ is the
dimensionality of $A$),  DeForcrand shows that
\begin{eqnarray}
P_{A}(\vec{\phi} \rightarrow \vec{\phi}') \approx {\rm erfc} (\epsilon\sqrt{\frac{N}{2}}).
\label{acc}
\end{eqnarray}
($\epsilon \equiv
\frac{||\vec{r}||}{||A\vec{\phi}'||}$) The computational overhead is simply one
matrix-vector product plus several matrix dot products per acceptance check. This can
save a factor of 2 to 3 in computer time. This is the general idea;
generalizations are also presented by DeForcrand.

\section{Accelerating Z(N) Noise}

Gaussian noise is not optimal for signal extraction\cite{liu}. Therefore
it is of interest to see if heatbath methods can be adapted to use
a mixture of Gaussian and Z(N) noise (Z(2) used here).

For this purpose, we begin with the expression,
\begin{eqnarray}
<{\cal O} > = \frac{1}{Z_{\phi}a^{2N}{\cal N}} \sum_{i} \int D \vec{\phi}
\,\, e^{-\frac{1}{a^{2}}|A\vec{\phi}-b\vec{Z}^{i}|^{2}} {\cal O},
\end{eqnarray}
where $\vec{Z}^{i}$ is a particular Z(2) noise vector and ${\cal N}= 2^{N}$ is the number of
Z(2) noises in the vector space and $a^{2}+b^{2}=1$. Let ${\cal
O} = \sum_{ij}\phi^{\dagger}_{j} A_{ji}^{\dagger} \phi_{i}$. One can do the integrals
to get,
\begin{eqnarray}
<\phi^{\dagger}_{j} A^{\dagger}_{ji} \phi_{i} > = 
\frac{1}{{\cal N}} \sum_{i}[ a^{2} \sum_{j} A^{-1}_{jj} + \nonumber \\
 b^{2} \sum_{j,k} Z^{i}_{j}A^{-1}_{jk}Z^{i}_{k}].
\end{eqnarray}
Then with 
\begin{eqnarray}
\frac{1}{{\cal N}}\sum_{i} Z^{i}_{j}Z^{i}_{k} = \delta_{jk},
\end{eqnarray}
we see that the answer is just ${\rm Tr} A^{-1}$, but with a weighting over Gaussian
($a^{2}$) and Z(2) ($b^{2}$) noises. We introduce
$\vec{\chi}$ as before,
\begin{eqnarray}
<{\cal O} > = \frac{1}{(\pi a^{4})^{N}Z_{\phi}{\cal N}} \int D \vec{\chi} \sum_{i} \int D
\vec{\phi} \nonumber \\
 e^{- \frac{1}{a^{2}}|A\vec{\phi}-b\vec{Z}^{i}|^{2} -
\frac{1}{a^{2}}|\vec{\chi}-A\vec{\phi}+b\vec{Z}^{i}|^{2}} {\cal O},
\end{eqnarray}
treating $\vec{Z}^{i}$ as a dynamical variable, the change in the action is now,
\begin{eqnarray}
\Delta S = \frac{1}{a^{2}} [|A\vec{\phi}'-b\vec{Z}^{i'}|^{2} +|\vec{\chi} -
A\vec{\phi}'+b\vec{Z}^{i'}|^{2}   \nonumber \\
 - |A\vec{\phi} -b\vec{Z}^{i}|^{2} - |\vec{\chi} - A\vec{\phi}+b\vec{Z}^{i} |^{2} ].
\end{eqnarray}
This is again a heatbath, with
\begin{eqnarray}
\Delta S =  2{\rm Re}(\vec{r^{a}}^{\dagger}\cdot (A^{a}\vec{\phi}- \frac{b}{a}\vec{Z}^{i} -
A^{a}\vec{\phi}'+\frac{b}{a}\vec{Z}^{i'})),
\label{this}
\end{eqnarray}
where $A^{a} \equiv A/a$ and $\vec{r}^{a} \equiv \vec{r}/a$. This is of the same form as
above and has the same acceptance since $(A^{a}\vec{\phi}- \frac{b}{a}\vec{Z}^{i})$ and
$(A^{a}\vec{\phi}'- \frac{b}{a}\vec{Z}^{i'})$ have variance $N$. The rescaled
$\vec{r}^{a}$ is used to define the residual, $\epsilon$, in the computer program.

\begin{figure}
\begin{center}
\epsfbox{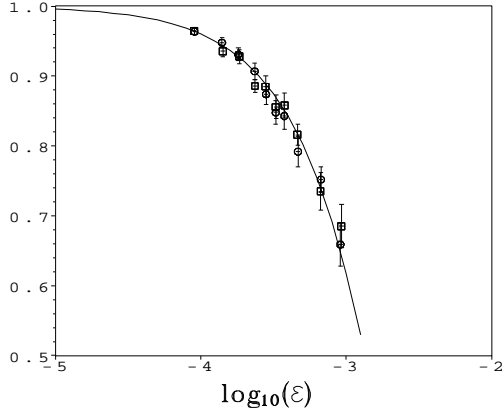}
\caption{The acceptance, $P_{A}$, as a function of 
${\rm log_{10}}(\epsilon)$ for $b/a=0$ (boxes) and $b/a=1$ 
(circles).\label{figure1}}
\end{center}
\vskip-1.cm
\end{figure}

\begin{figure}
\begin{center}
\epsfbox{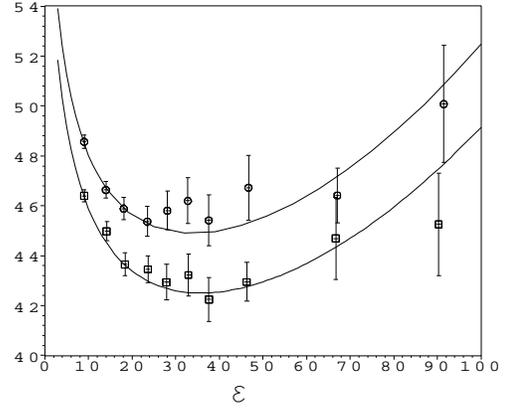}
\caption{The number of mixed Z(2)/Gaussian heatbath iterations divided
by the model acceptance, $N_{mixed}/P_{A}$, as a function of $\epsilon$
in units of $10^{-5}$ for $b/a=0$ (boxes) and $b/a=1$ 
(circles).\label{figure2}}
\end{center}
\vskip-1.cm
\end{figure}

Although the acceptance is the same, the number of
iterations is greater for a given cutoff, $\epsilon$, on the 
new residual vector
since it is defined by dividing by $a<1$. However, 
since the convergence on
the residual is typically exponential, changes in $a$ are 
accomodated by a modest number
of extra iterations. The fact that the acceptance is the same 
at the rescaled $\epsilon$ value we will see has the helpful
consequence that one does not have to re-search for the optimum 
$\epsilon$ value at which to
run, even though now we use a noise mixture.

The above discussion can be generalized to justify more complicated 
exponential shifts
$A\vec{\phi}\rightarrow A\vec{\phi}-\vec{C}^{i}$ where $\vec{C}^{i}$ 
can depend on other
noises, parameters, etc., besides the specific one $\vec{C}^{i} = 
b\vec{Z}^{i}$ used
above. I am using a solver which has enforced even/odd preconditioning in the
Wilson/Dirac matrix $M$. One can show that in this case one is not 
simulating $A=M$ directly as
in the above discussion, but a different, shifted system and the
residual vector is purely in one sector (even/odd) or the other.

Figs.~1 and 2 show acceptance data on a $12^{3}\times 24$ lattice using the
Wilson/Dirac matrix at $\kappa = 0.148$. I am calculating the average 
acceptance on 100 noises at various cutoff values of $\epsilon$ and with 
$b/a=0$ and 1. For this solver I get about a
$6\%$ increase in the number of iterations for this ratio. (The model 
suggests about
$14\%$ increase in the number of iterations for an $80\%$ mixture ($b/a=2$) 
and about a
$30\%$ increase for a $97.3\%$ mixture ($b/a=6$) for the parameters in 
this simulation.) 
Fig.\ 1 shows $P_{A}$ in the crucial window of $\epsilon$ of from
$10^{-4}$ through $10^{-3}$, where it varies between $\sim 0.97$ to $0.65$
for $b/a=0,1$ for a single gauge configuration. The acceptance is the 
same for both $b/a$ values within errors and agrees with Eq.(\ref{acc}). The
reason this interval is crucial is illustrated in Fig.\ 2. There we see
that the number of mixed Z(2)/Gaussian heatbath iterations divided 
by the acceptance, 
$N_{mixed}/P_{A}$, has a minimum at $\epsilon \sim 4\times 10^{-4}$
for $b/a=0,1$. Shown here also is the model value for this quantity given by
\begin{equation}
N_{mixed}/P_{A} = \frac{-\frac{1}{c} {\rm ln}(a\epsilon \frac{\sqrt{N}}{r_{0}})}
{{\rm erfc}(\epsilon \sqrt{\frac{N}{2}})},\label{num}
\end{equation}
where $N_{mixed}$ is assumed determined by $||\vec{r}|| = r_{0}e^{-cN_{mixed}}$.
As noted above, I am using a solver which has enforced even/odd 
preconditioning in the
Wilson/Dirac matrix $M$. The upshot for this simulation is that $N$ 
in (\ref{acc}) and (\ref{num}) must 
include a factor of $\frac{1}{2}$, $N=\frac{1}{2}\times (12^{3}\times 
24\times 4\times 3)$.

Fig.3 shows the normed effective mixed Z(2)/Gaussian
variance in the operator $\bar{\psi}\psi$ as a function of $b/a$. 
It has a very shallow minimum at $b/a \approx 10$.
By normed effective variance I mean the ratio $(V_{mixed}\times 
N_{mixed})/(V_{Gauss}\times N_{Gauss})$, which takes into account
that $N_{mixed}>N_{Gauss}$ for $b \ne 0$. The model
and data suggest that the minimum of this ratio is $\approx 0.3$.
The horizontal line gives the fully converged value of the variance ratio
$V_{Z(2)}/V_{Gauss}$, which is $\approx 0.2$. Thus, one looses a factor
of $\approx 2/3$ in the effective variance compared to the fully
converged Z(2) simulation. The gain in computer time is reduced from
a factor of from 2 to 3 to a factor of from 1.33 to 2. These numbers
are apparently typical. Other operators with larger converged
$V_{Z(2)}/V_{Gauss}$ ratios have sharper minima at smaller $b/a$.

\section{Conclusions and Observations}

Ref.\cite{phil} shows that heatbath methods can speed up simulations of 
many disconnected loop operators or by a factor of $\approx 2$ to 3.
However, Gaussian noise is not optimal and so heatbath methods do not help 
for operators whose variance is diagonally dominate, such as Wilson
$\bar{\psi}\psi$. It has been shown here that these methods 
can be generalized to a mixture of Gaussian and 
Z(N) noise. With an mixture/iteration penalty factor of about $2/3$,
the Gaussian noise accelerates the Z(N) sector.

For diagonally dominate operators, there exists
a Z(N)/Gaussian ratio that minimizes the variance. The noise can then be tailored to
the operator (\lq\lq designer noise"). The optimum $b/a$ parameter can be numerically 
estimated from the parameters of the model and an 
independent measurement of the fully converged Z(N)/Gaussian variance ratio.

It is not shown here, but the even/odd structure of the Wilson/Dirac
matrix can be exploited to increase the computer time gain of 
diagonally dominate operators further by restricting the Gaussian noise to only one sector
or the other. This and other aspects of heatbath noise methods will 
be discussed in a future publication.

\begin{figure}
\begin{center}
\epsfbox{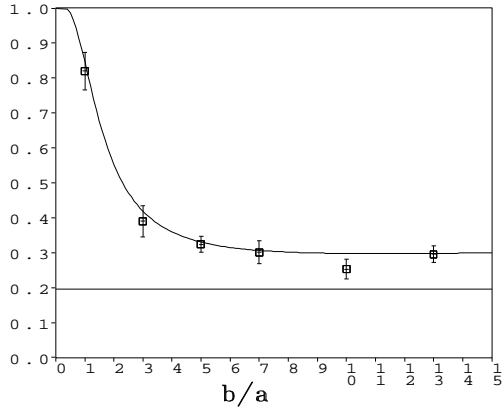}
\caption{Effective variance of the $b \ne 0$ simulation for $\bar{\psi}\psi$ normalized
to the Gaussian simulation with $b=0$ as a function of $b/a$.
\label{figure3}}
\end{center}
\vskip-1.cm
\end{figure}

\section{Acknowledgements}

This work was supported by NSF Grant No.\ 0070836 and
the Baylor University Sabbatical program. The calculations were done at 
NCSA and utilized the SGI Origin 2000 System at the University of 
Illinois. The author thanks P.\ de Forcrand for helpful comments.

\end{document}